\documentclass[aps,pra,twocolumn,amsmath,amssymb,nofootinbib,showpacs,superscriptaddress]{revtex4-1}

\usepackage[english]{babel}
\usepackage{latexsym}
\usepackage{graphics}
\usepackage{graphicx}
\usepackage{epsfig}
\usepackage{color}
\usepackage{bm}
\usepackage{amsmath}
\usepackage{amssymb}
\usepackage{amsthm}
\usepackage{dcolumn}
\usepackage{bm}
\usepackage{float}
\usepackage{hyperref}
\usepackage{epstopdf}
\usepackage{cleveref}
\usepackage[svgnames]{xcolor}
\usepackage{enumerate}
\hypersetup{hidelinks,colorlinks=true,allcolors=DarkBlue}

\usepackage{lineno}

\begin{document}
	
	\preprint{APS/123-QED}
	
	\title{Tomographic and entropic analysis of modulated signals}
	
	\author{A.S. Mastiukova}
	\affiliation{Russian Quantum Center, Skolkovo, Moscow 143025, Russia}
	\affiliation{Moscow Institute of Physics and Technology, Dolgoprudny, Moscow Region 141700, Russia} 
	
	\author{M.A. Gavreev}
	\affiliation{Russian Quantum Center, Skolkovo, Moscow 143025, Russia}
	\affiliation{Moscow Institute of Physics and Technology, Dolgoprudny, Moscow Region 141700, Russia} 
	
	\author{E.O. Kiktenko}
	\affiliation{Russian Quantum Center, Skolkovo, Moscow 143025, Russia}
	\affiliation{Moscow Institute of Physics and Technology, Dolgoprudny, Moscow Region 141700, Russia} 
	\affiliation{Department of Mathematical Methods for Quantum Technologies, Steklov Mathematical Institute of Russian Academy of Sciences, Moscow 119991, Russia}
	
	\author{A.K. Fedorov}
	\affiliation{Russian Quantum Center, Skolkovo, Moscow 143025, Russia}
	\affiliation{Moscow Institute of Physics and Technology, Dolgoprudny, Moscow Region 141700, Russia} 
	
	\date{\today}
	\begin{abstract}
		We study an application of the quantum tomography framework for the time-frequency analysis of modulated signals. 
		In particular, we calculate optical tomographic representations and Wigner-Ville distributions for signals with amplitude and frequency modulations.  
		We also consider time-frequency entropic relations for modulated signals, which are naturally associated with the Fourier analysis. 
		A numerical toolbox for calculating optical time-frequency tomograms based on pseudo Wigner-Ville distributions for modulated signals is provided.
	\end{abstract}
	
	\maketitle
	
\section{Introduction}
	
Time-frequency analysis is a powerful tool of modern signal processing~\cite{Cohen1995,Papandreou2002,Dragoman2005,Sejdic2009}.
Complementary to the information that can be extracted from the frequency domain via Fourier analysis, time-frequency analysis provides a way for studying a signal in both time and frequency representations simultaneously.
This is useful, in particular, for signals of a sophisticated structure that change significantly over their duration, for example, music signals~\cite{Pielemeier1996}.
Existing approaches to time-frequency analysis use linear canonical transformations preserving the symplectic form~\cite{Cohen1995}.
Geometrically this can be illustrated as follows: 
the Fourier transform can be viewed as a $\pi/2$ rotation in the associated time-frequency plane, whereas other time-frequency representations allow arbitrary symplectic transformations in the time-frequency plane. 
There is a number of ways for defining a time-frequency distribution function with required properties (for a review, see Ref.~\cite{Sejdic2009}). 
Transformations between various distributions in time-frequency analysis are quite well-understood~\cite{Cohen1995}.
	
The idea behind time-frequency analysis is very close to the motivation for studying phase-space representations in quantum physics. 
As it is well known, the relation between position and momentum representations of the wave function is given by the Fourier transform, which is similar to the relation between signals in time and frequency domains. 
This analogy becomes even more transparent in the framework of analytic signals, which are complex as well as wave functions.
One of the possible ways to characterize a quantum state in the phase space are to use the Wigner quasiprobability distribution~\cite{Wigner1932}.
The Wigner quasiprobability distribution resembles classical phase space probability distributions that is used in statistical mechanics. 
However, it cannot be fully interpreted as a probability distribution since it takes negative values~\cite{Wigner1932,Wigner1984,Ferry2018}.
We also note the Wigner quasiprobability distribution is successfully used for analyzing various phenomena in quantum optics and quantum statistical physics~\cite{Ferry2018}.
In the field of signal processing, the Wigner distribution function often referred to as the Wigner--Ville distribution~\cite{Ville1948,Boashash1988}.
The application of the Wigner distribution makes an interesting connection between the methods in quantum physics and signal processing,
especially in the context of time-frequency and position-momentum uncertainty relations. 
	
Recent decades, the link between phase space formulation of quantum mechanics and time-frequency analysis intensively studied in the context of quantum tomography~\cite{Manko1996}.
Quantum tomography appears as a technique for the reconstruction of the Wigner function (density matrix) in quantum-optical experiments~\cite{Lvovsky2009}.
The results of tomographic measurements in principle contain all the information about the measured system, so they can be considered as a quantities for the description of quantum states~\cite{Manko1996}. 
This is the core idea behind the omographic representation of quantum states. 
In particular, symplectic tomography protocols use a marginal probability distribution of shifted and squeezed position and momentum variables.
This approach has been used in the context of time-frequency analysis~\cite{Manko1999} and time-frequency entropic analysis~\cite{MankoMA2006} for various types of signals, 
such as complex Gaussian signals~\cite{Manko1999,MankoMA2006,Mendes2001,Manko2005,Manko2012,MankoMA2000} and reflectometry data~\cite{Mendes2009,Mendes2015}.
A general analysis of the relation between time-frequency tomograms and other transformations (including wavelets) is presented in Ref.~\cite{Mendes2001}.
However, the considered examples of signals lack analyzing modulated signals, which are intensively used in telecommunication signals.
Moreover, the Wigner-Ville distribution has been considered in the context of diagnostics of features of modulated signals~\cite{Trajin2015},
so one can expect that tomograms are helpful for such an analysis.
	
In this work, we consider tomographic representations for modulated signals.
We calculate optical tomographic representations and Wigner-Ville distributions for signals with amplitude and frequency modulations.  
In particular, we study the method of the optical time-frequency tomograms via pseudo Wigner-Ville distributions, and discuss advantages of such an approach.
We also consider time-frequency entropic relations for modulated signals. 

\begin{figure*}[htbp]
	\includegraphics[width=0.87\linewidth]{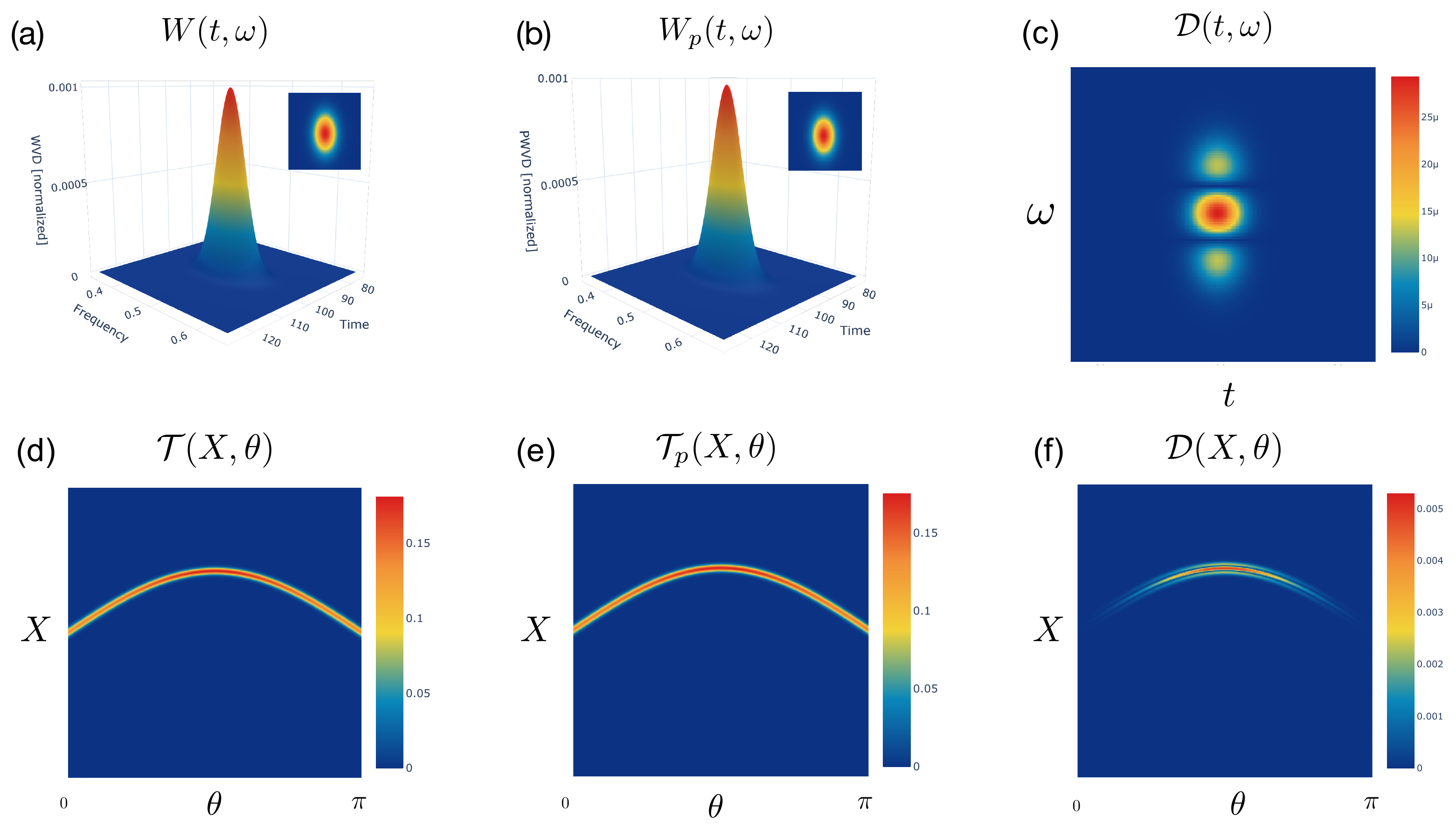}
	\vskip -4mm
	\caption{time-frequency analysis for the chirp signal for the following set of parameters ($A\approx{0.166}$, $\phi_0=0$, $\alpha=0.03$, $t_0=100$, and $\omega=\pi$) given by Eq.~(\ref{chirp}):
		(a) the Wigner-Ville distribution is presented (the 2D representation in the time-frequency plane is in insert);
		(b) the pseudo Wigner-Ville distribution is presented (the 2D representation in the time-frequency plane is in insert);
		(c) the difference between Wigner-Ville and pseudo Wigner-Ville distributions is illustrated;
		(d) the optical time-frequency tomogram is presented;
		(e) the optical time-frequency tomogram based on the calculation of the pseudo Wigner-Ville distributions is presented;
		(f) the difference between tomograms is illustrated.} 
	\label{fig:chirp}
\end{figure*}
	
Our work is organized as follows. 
In Sec.~\ref{sec:general}, we introduce general relations for tomographic analysis of analytic signals.
In Sec.~\ref{sec:modulated}, we calculate optical tomographic representations for signals with amplitude and frequency modulations.
In Sec.~\ref{sec:entropic}, we analyze time-frequency entropic relations.
We conclude in Sec.~\ref{sec:conclusion}.
		
\section{Tomographic analysis}\label{sec:general}
	
Here we introduce basic tools for the tomographic analysis of signals. 
We consider a time-dependent signal $s(t)$, whose representation in the frequency domain $\tilde{s}(\omega)$ can be obtained via the Fourier transform.
Conventionally, we use an analytical representation of signals in the following form:
\begin{equation}\label{analyt}
	\mathcal{S}(t)=s(t)+iH\left[s(t)\right],
\end{equation}
where
\begin{equation}
	H\left[s(t)\right]=\frac{1}{\pi}\int_\mathbb{R}{dp\frac{s(p)}{t-p}}
\end{equation}
is the Hilbert transform of the signal.
The advantage of using the analytic signal is that in the frequency domain the amplitude of negative frequency components are zero. 
This satisfies mathematical completeness of the problem by accounting for all frequencies, yet does not limit the practical application since only positive frequency components have a practical interpretation.
The method based on the use of analytic signals also makes a clear analogy between time-frequency distributions in signal processing and phase-space distributions in quantum mechanics~\cite{Cohen1995}.
	
The family of marginal distributions, which contains complete information on the analytical signal, has been introduced in Ref.~\cite{Manko1996}.
It has the following form:
\begin{equation}\label{tomogram}	
	\mathcal{T}(X,\theta)=\frac{1}{2\pi|\sin\theta|}\left|\mathcal{I}(X,\theta)\right|^2,
\end{equation}
where
\begin{equation}\label{eq:fft}
	\mathcal{I}(X,\theta)=\int_\mathbb{R}{dt\mathcal{S}(t)\exp\left[\frac{it^2\cos\theta}{2\sin\theta}-\frac{itX}{\sin\theta}\right]}.
\end{equation}
Here $X={t}\cos\theta+{\omega}\sin\theta$ is the dimensionless quadrature variable.
This representation is referred to as the optical time-frequency tomogram of the signal $\mathcal{S}(t)$.
The integral transformation in Eq.~(\ref{eq:fft}) is the fractional Fourier transform.
The tomogram is normalized as follows:
\begin{equation}
	\int_\mathbb{R}{dX\mathcal{T}(X,\theta)}=1.
\end{equation}
It also gives the distribution of the signal in time and frequency domains, correspondingly:
\begin{equation}
	\mathcal{T}(X=t,0)=|\mathcal{S}(t)|^2, \quad \mathcal{T}(X=\omega,\pi/2)=|\tilde{\mathcal{S}}(\omega)|^2.
\end{equation}
The optical time-frequency tomogram is a particular case of the symplectic time-frequency tomogram $T(X,\mu,\nu)$ of the signal $\mathcal{S}(t)$,
where $\mu=\cos\theta$ and $\nu=\sin\theta$.
In some cases, another variations of tomographic representation, such as time-scale tomograms, frequency-scale tomograms, and time-conformal tomograms, are used~\cite{Mendes2015}.
We restrict ourselves to the consideration of optical time-frequency tomograms only.
	
It seems to be quite straightforward to calculate optical time-frequency tomograms using Eq.~(\ref{tomogram}).
However, there are well-known problems in the field of signal processing, such as, for example, aliasing, which give rise to distortions during the signal reconstruction and computational difficulties. 
These problems also occur during the calculation of the Wigner-Ville distribution for analytic signals~\cite{Boashash1988}. 
We remind that the Wigner-Ville distribution of the signal has the following form:
\begin{equation}\label{Wign}
	W(t,\omega)=\int_\mathbb{R}{d\tau\mathcal{S}\left(t+\frac{\tau}{2}\right)\mathcal{S}^*\left(t-\frac{\tau}{2}\right)e^{-i\omega \tau}}.
\end{equation}
The Wigner-Ville distribution is normalized as follows:
\begin{equation}\label{Wign}
	\frac{1}{2\pi}\int_{\mathbb{R}^2}{dtd\omega W(t,\omega)}=1.
\end{equation}
When the Wigner-Ville distribution is applied to a signal with multi frequency components, cross-terms appear due to its quadratic nature.
In order to avoid the effect of cross-terms the windowed version of the Wigner-Ville distribution, which is known as pseudo Wigner-Ville distribution, is used. 
The pseudo Wigner-Ville distribution is defined as follows:
\begin{equation}\label{pWign}
	W_{p}(t,\omega)=\int_\mathbb{R}{d\tau h(\tau)\mathcal{S}\left(t+\frac{\tau}{2}\right)\mathcal{S}^*\left(t-\frac{\tau}{2}\right)e^{-i\omega \tau}},
\end{equation}
where $h(\tau)$ is the window function in the time domain.
The window function $h(\tau)$ can be used, for example, in the Hamming window form:
\begin{equation}
	h_{M}(\tau){=}0.54-0.46\cos\left(\frac{2\pi\tau}{M-1}\right), \, 0\leq\tau\leq{M-1}.
\end{equation}
	
\begin{figure}[t]
\begin{centering}
	\includegraphics[width=0.965\columnwidth]{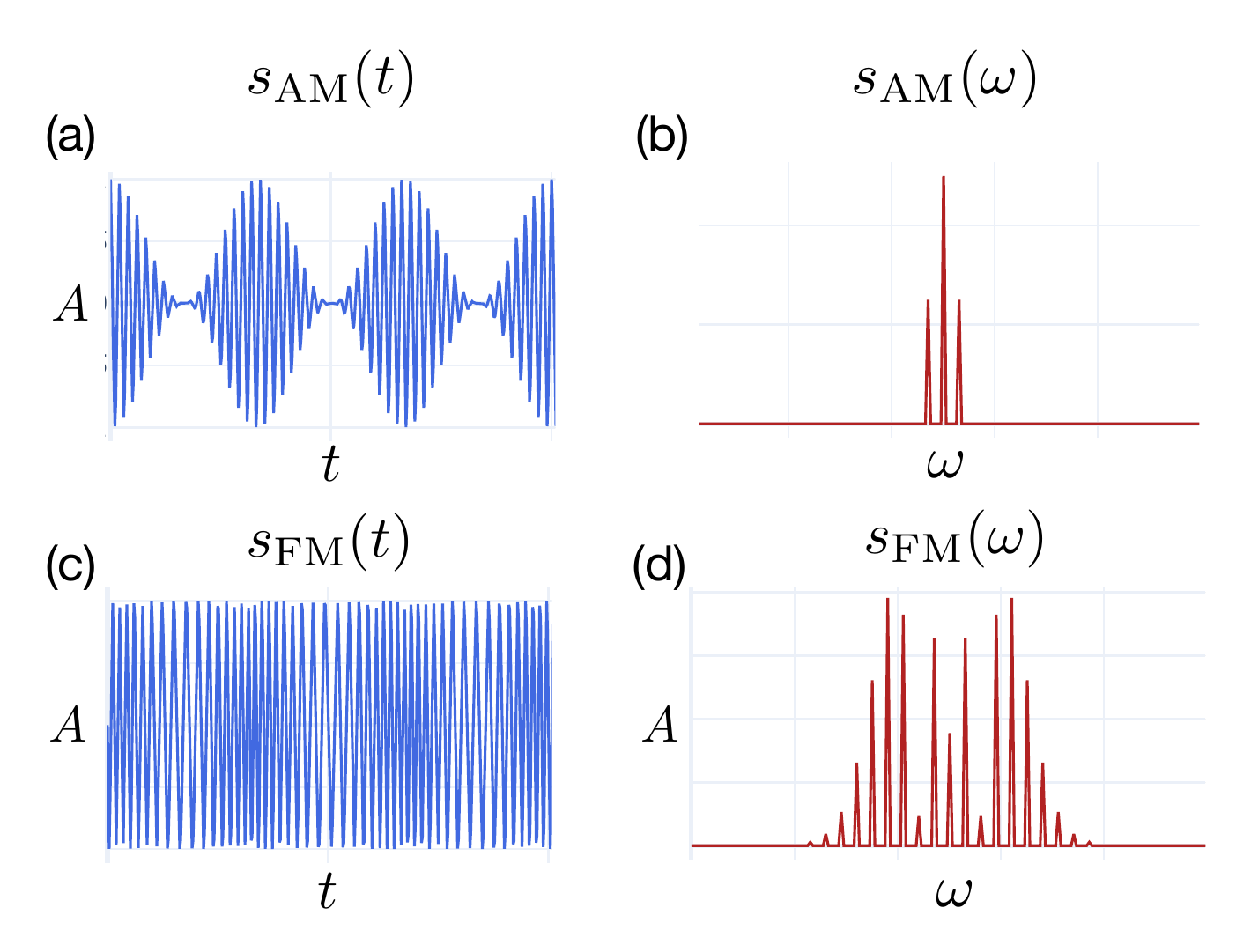}
	\end{centering}
	\vskip -4mm
	\caption
	{Considered modulated signals: 
	(a) AM signal in the time domain and (b) in the frequency domain; 
	(c) FM signal in the time domain and (d) in the frequency domain.}
\label{fig:AMFM}
\end{figure}
	
Pseudo Wigner-Ville distributions~\cite{Shin1993,Katkovnik1998,Katkovnik19982,Sucic2009} and their modifications are actively used in various fields, 
such as the dispersion analysis of waveguides~\cite{Jukam2016} and studying oil-in-water flow patterns~\cite{{Jin2016}}.
We note that various time-frequency filters are employed for eliminating cross-terms in the Wigner-Ville distribution, which is of high importance for the analysis of non-stationary systems.
Another approach for reducing cross-terms in the Wigner-Ville distribution uses a tunable-Q wavelet transform~\cite{Nishad2016}.
In our consideration below, we use the simplest case of the pseudo Wigner-Ville distribution with the simplest form of the window function.
	
Using the relation between Wigner-Ville distributions and optical time-frequency tomograms, which is given by the Radon transform, one can reconstruct the optical tomogram of the signal as follows:
\begin{equation}	
	\!\!\!\!\!\mathcal{T}(X,\theta)=\int_{\mathbb{R}^3}{\frac{dkdtd\omega}{(2\pi)^2}W(t,\omega)e^{-ik(X-t\cos\theta-\omega\sin\theta)}}.
\end{equation}
Then in order to reduce the complexity of calculating optical time-frequency tomograms it is possible to redefine it via pseudo Wigner-Ville distributions as follows:
\begin{equation}	
	\!\!\!\!\!\mathcal{T}_{p}(X,\theta)=\int_{\mathbb{R}^3}{\frac{dkdtd\omega}{(2\pi)^2}W_{p}(t,\omega)e^{-ik(X-t\cos\theta-\omega\sin\theta)}}.
\end{equation}

This relation give rise to the modification of the integral relation between analytic signal and its time-frequency optical tomogram, which is given by Eq.~(\ref{tomogram}).
In our work, we use the pseudo time-frequency optical tomogram $\mathcal{T}_{p}(X,\theta)$ for analyzing properties of signals.
We note that this consideration is related to the establishing a correspondence between the fractional Fourier transform and the Wigner distribution~\cite{Mustard2006,Lima2017}.
	
In order to see a difference in calculating original and pseudo time-frequency optical tomograms, we consider an example of a chirp signal of the following form:
\begin{equation}\label{chirp}
	s(t)=A\sin\left(\omega t +\phi_0\right)e^{-\alpha\left(t-t_0\right)^2},
\end{equation}
where $A$ is the fixed amplitude, $\phi_0$, $\alpha$, and $t_0$ are fixed constants.
For this chirp signal in the analytic form given by Eq.~(\ref{analyt}) we calculate first original Wigner-Ville distribution (Fig.~\ref{fig:chirp}a) and pseudo Wigner-Ville distribution (Fig.~\ref{fig:chirp}b).
One can capture a difference between $\mathcal{D}(t,\omega)=|W(t,\omega)-W_p(t,\omega)|$ original Wigner-Ville distribution and pseudo Wigner-Ville distribution (see Fig.~\ref{fig:chirp}c).
This difference manifests in calculating original and pseudo time-frequency optical tomograms (Fig.~\ref{fig:chirp}d, Fig.~\ref{fig:chirp}e, and Fig.~\ref{fig:chirp}f),
where $\mathcal{D}(X,\theta)=|\mathcal{T}(X,\theta)-\mathcal{T}_p(X,\theta)|$.
The differences $\mathcal{D}(t,\omega)$ and $\mathcal{D}(X,\theta)$ are non-zero.
This can be a signature of the fact that the signal can be sensitive to the presence of the time-window, which is of importance for capturing properties of non-stationary signals.

\section{Tomographic representation for modulated signals}\label{sec:modulated}
	
\begin{figure}[t]
	
\begin{centering}
	\includegraphics[width=1\columnwidth]{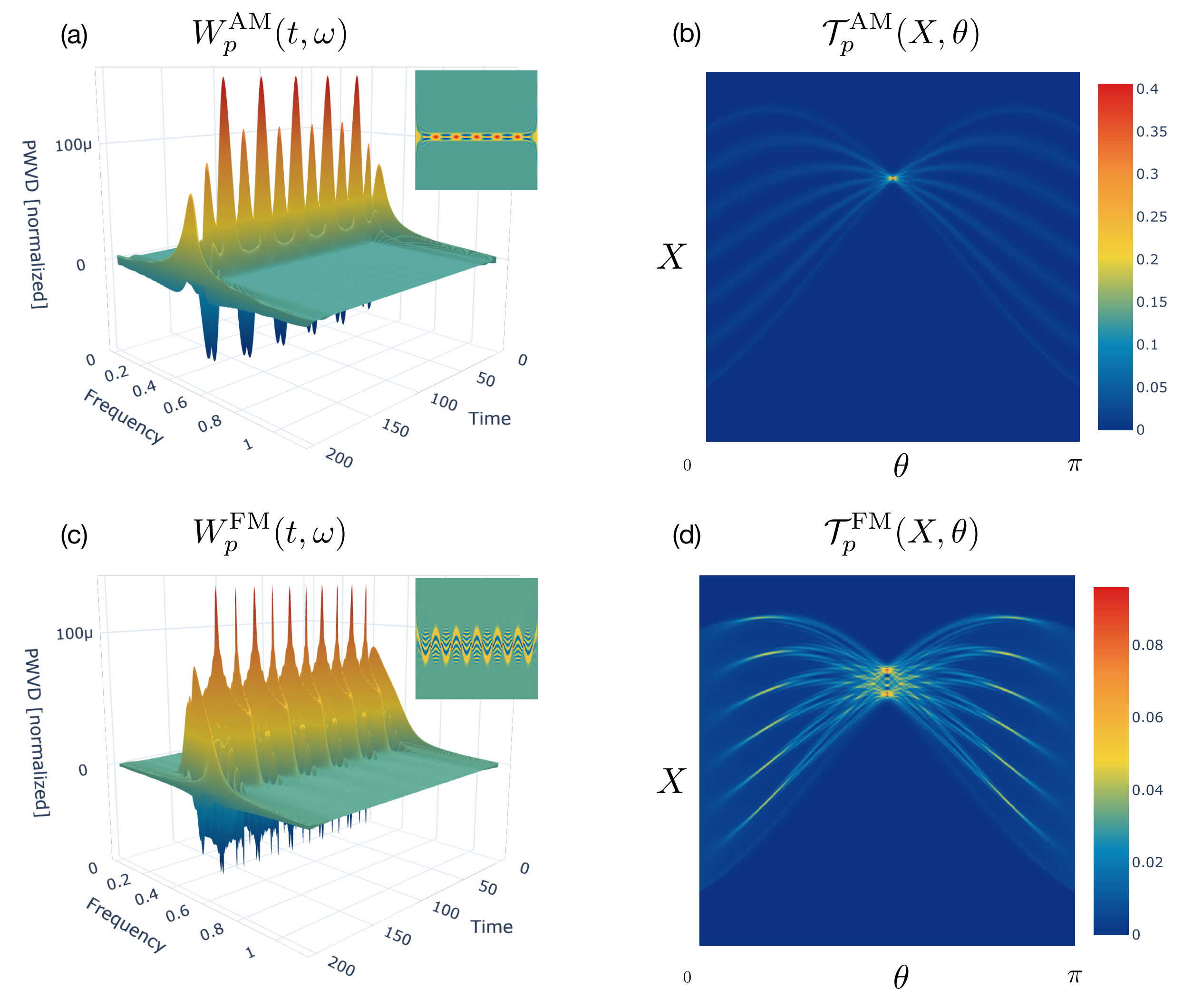}
	\end{centering}
	\vskip -4mm
	\caption
	{Results for modulated signals: (a) Pseudo Wigner-Ville distribution and (b) optical time-frequency tomograms for AM signals are presented;
	(a) Pseudo Wigner-Ville distribution and (b) optical time-frequency tomograms for FM signals are illustrated.}
	\label{fig:Tomograms}
\end{figure}
	
Several types of signals have been considered in the context of tomographic analysis~\cite{Manko1999,MankoMA2006,Mendes2001,Manko2005,Manko2012,MankoMA2000}.
However, existing examples lack of analyzing modulated signals, which are intensively used in telecommunication tasks.
A general model of signals that we are interested in has the following form:
\begin{equation}
	s(t) = A(t)\cos(\omega(t)t + \phi_0),
\end{equation}
where $A(t)$ is the amplitude of the signal, $\omega(t)$ is the frequency and $\phi_0$ is the phase. 
The fact that the amplitude and frequency are time-dependent indicates that they can be used for modulation purposes. 
Most radio systems in the 20th century used frequency modulation (FM) or amplitude modulation (AM) for radio broadcast.

We start from the simplest case of signals with AM.
A signal with amplitude modulation is as follows:
\begin{equation}\label{eq:AM}
	s_{\rm AM}(t) = A(t)\cos(\omega t + \phi_0),
\end{equation}
where $A(t)$ is a law of change of amplitude with time.
In amplitude modulation, the amplitude (signal strength) of the carrier wave is varied in proportion to that of the message signal being transmitted.
We consider the following case:
\begin{equation} 
	A(t)=1+ms_m(t), \quad s_m(t)=\cos(\Omega t),
\end{equation}
where $m$ is the modulation coefficient and $\Omega$ is the frequency of the signal.
We illustrate modulated signals and their Fourier transforms in Fig.~\ref{fig:AMFM}a and Fig.~\ref{fig:AMFM}b.
	
Another case is to consider a signal with FM, which has the following form:
\begin{equation} \label{eq:FM}
	s_{\rm FM}(t) = A\cos\left(\omega_0t+ \omega_d\int_\mathbb{R}{dt s_m(t)} + \phi_0\right).
\end{equation}
In this case, the frequency varies as follows:
\begin{equation}
	\omega(t)=\omega_0+\omega_ds_m(t),
\end{equation}
where $\omega_d$ is frequency deviation, i.e. an analog of the modulation parameter $m$ for the amplitude modulation signal.
We illustrate the signal and its Fourier transform in Fig.~\ref{fig:AMFM}c and Fig.~\ref{fig:AMFM}d.
	
For the signals with AM and FM given by Eq.~(\ref{eq:AM}) and Eq.~(\ref{eq:FM}), correspondingly, we calculate modified optical time-frequency tomograms based on pseudo Wigner-Ville distributions. 
These results are presented in Fig.~\ref{fig:Tomograms}.

\begin{figure}[t]
	\begin{centering}
	\includegraphics[width=1\columnwidth]{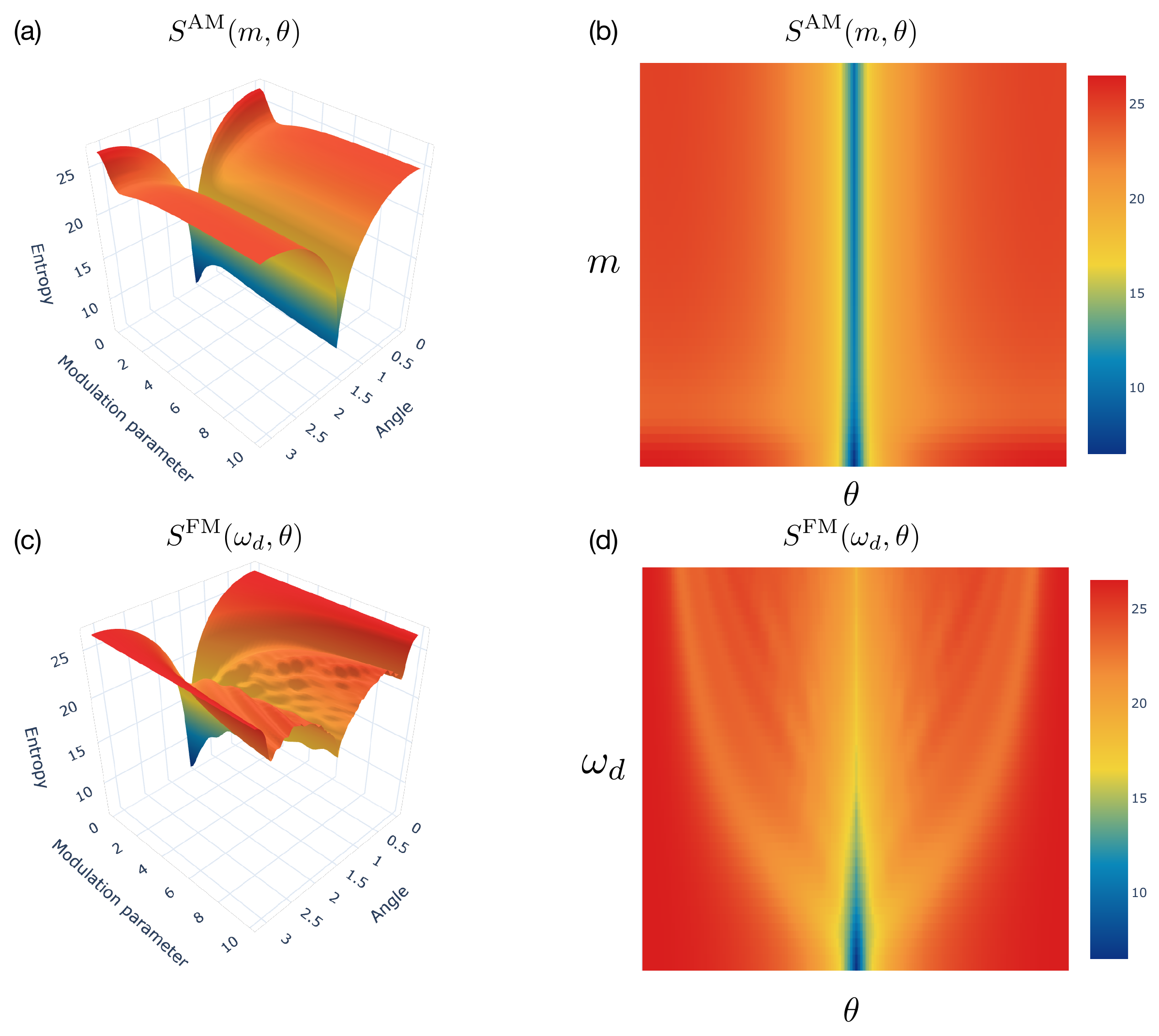}
	\end{centering}
	\vskip -4mm
	\caption
	{Entropies as function of the angle $\theta$ and modulation parameter for (a) AM analytic signal and (b) FM analytic signal.}
	\label{fig:entropy}
\end{figure}

\section{Entropic relations}\label{sec:entropic}

Another interesting point of view on the link between signal processing and quantum physics originates from uncertainty relations and related entropic relations.
The idea behind this consideration is the fact that for analytic signals with normalized energy of the spectrum,
\begin{equation}
	\int_\mathbb{R}{dt|\mathcal{S}(t)|^2}=\int_\mathbb{R}{d\omega|\tilde{\mathcal{S}}(\omega)|^2}=1,
\end{equation}
one can think of the introduction of the differential entropy (also known as the continuous Shannon entropy) in the time domain as follows:
\begin{equation}
	S_t =-\int_\mathbb{R}{dt|\mathcal{S}(t)|^2 \ln{|\mathcal{S}(t)|^2}}.
\end{equation}
The differential entropy on the analytic signal in the frequency domain has the following form:
\begin{equation}
	S_\omega=-\int_\mathbb{R}{d\omega |\tilde{\mathcal{S}}(\omega)|^2 \ln{|\tilde{\mathcal{S}}(\omega)|^2}}.
\end{equation}
One can see that these expressions for differential entropies are equivalent to those for the position $|\psi(q)|^2$ and momentum $|\psi(p)|^2$ representations of the probability distribution function, which are calculated via the corresponding wavefunction. 
Therefore, there the following entropic inequality holds for the differential entropies of analytic signals~\cite{Mycielski1975}:
\begin{equation} \label{uncert}
	S_t+S_\omega\geq\mathrm{ln}(\pi e).
\end{equation}

The considered below tomographic approach to signal analysis allows introducing the time-frequency entropy as follows:
\begin{equation}\label{uncert}
	S(\theta)=-\int_\mathbb{R}{dX\mathcal{T}_{p}(X,\theta)\ln{\mathcal{T}_{p}(X,\theta)}}.
\end{equation}
We note that the optical time-frequency tomogram $\mathcal{T}_{p}(X,\theta)$ is calculated on the basis of the pseudo Wigner-Ville distribution. 

We use this formula for the analysis of AM and FM signals. 
In this case, $S(\theta)$ also becomes a function of the modulation parameter, so we have $S^{\rm AM}(\theta,m)$ and $S^{\rm FM}(\theta,\omega_d)$ for signals given by Eq.~(\ref{eq:AM}) and Eq.~(\ref{eq:FM}), correspondingly. 
We present the results of calculations time-frequency entropies based on optical time-frequency tomograms in Fig.~\ref{fig:entropy}.
We also can check entropic relations given by Eq.~(\ref{uncert}) for tomograms formulated as follows:
\begin{equation}
	S(\theta)+S(\theta+\pi/2)\geq\mathrm{ln}(\pi e).
\end{equation}

\section{Conclusion} \label{sec:conclusion}
	
We have considered applications of quantum tomography framework for the time-frequency analysis of signals with amplitude and frequency modulations. 
We have demonstrated an efficient way for calculating optical time-frequency tomograms for analytic signals based on the pseudo Wigner-Ville distribution, 
which seems to be important for signals of a sophisticated structure that change significantly over their duration.
We also have analyzed differential entropies of the signals calculated via optical time-frequency tomograms and discussed corresponding entropic relations.

Our approach can be extended and generalized in a number of ways.
First, one can think of studying time-frequency (symplectic or optical) tomograms based on other types of modified Wigner-Ville distributions, such Wigner-Ville distributions with windows both in time and frequency domains. 
Second, an important is to understand the nature of the difference between original time-frequency tomograms and modified time-frequency tomograms.
In particular, it is important whereas modified time-frequency tomograms are able to capture some feature of highly non-stationary signals that are important. 
Finally, an interesting task to analyse how time-frequency tomograms can be measured in various applications, such as analysis of reflectometry data~\cite{Mendes2009,Mendes2015}.

\medskip
\section*{Acknowledgments}
The work was supported by the grant of the President of the Russian Federation (project MK- 923.2019.2).


\begin{thebibliography}{99}
		
	\bibitem{Cohen1995}
	L. Cohen,	
	{\it time-frequency Analysis}
	(Prentice-Hall, New York, 1995).
		
	\bibitem{Papandreou2002}
	A. Papandreou-Suppappola,	
	{\it Applications in time-frequency Signal Processing}
	(CRC Press, Boca Raton, 2002).

    	\bibitem{Dragoman2005}	
	D. Dragoman,
	Applications of the Wigner distribution function in signal processing,
	{\href{https://doi.org/10.1155/ASP.2005.1520}{EURASIP J. Adv. Signal Process. {\bf 10}, 1520 (2005)}}.
	
	\bibitem{Sejdic2009}		
	E. Sejdi\'{c}, I. Djurovi\'{c}, and J. Jiang,
	Time-frequency feature representation using energy concentration: An overview of recent advances,
 	{\href{https://doi.org/10.1016/j.dsp.2007.12.004}{Digit. Signal Process. {\bf 19}, 153 (2009)}}.

	\bibitem{Pielemeier1996}		
	W.J. Pielemeier, G.H. Wakefield, and M.H. Simoni, 
	time-frequency analysis of musical signals,
	{\href{http://dx.doi.org/10.1109/5.535242}{IEEE Proc. {\bf 84}, 1216 (1996)}}.
		
	\bibitem{Wigner1932}
	E.P. Wigner,
	On the quantum correction for thermodynamic equilibrium,
	{\href{http://dx.doi.org/10.1103/PhysRev.40.749}{Phys. Rev. {\bf 40}, 749 (1932)}}.	
	
	\bibitem{Wigner1984}
	M. Hillery, R. OConnell, M. Scully, and E. Wigner,
	Distribution functions in physics: Fundamentals,
	{\href{http://dx.doi.org/10.1016/0370-1573(84)90160-1}{Phys. Rep. {\bf 106}, 121 (1984)}}.
	
	\bibitem{Ferry2018}
	J. Weinbub and D.K. Ferry,
	Recent advances in Wigner function approaches,
	{\href{http://dx.doi.org/10.1063/1.5046663}{Appl. Phys. Rev. {\bf 5}, 041104 (2018)}}.
	
	\bibitem{Ville1948}
	J. Ville, 
	Th\'eorie et applications de la notion de signal analytique, 
	Cable Transm. {\bf 2}, 61 (1948).
	
	\bibitem{Boashash1988}				
	B. Boashash,
	Note on the use of the Wigner distribution for time frequency signal analysis,
	{\href{https://doi.org/10.1109/29.90380}{IEEE Trans. on Acoust. Speech. and Signal Processing {\bf 36}, 1518 (1988)}}.	
	
	\bibitem{Manko1996}
	S. Mancini, V. I. Man’ko, P. Tombesi, 
	Symplectic tomography as classical approach to quantum systems,
	{\href{http://dx.doi.org/10.1063/1.5046663}{Phys. Lett. A {\bf 213}, 1 (1996)}}.
	
	\bibitem{Lvovsky2009}
	A.I. Lvovsky and M.G. Raymer, 
	Continuous-variable optical quantum-state tomography,
	{\href{https://doi.org/10.1103/RevModPhys.81.299}{Rev. Mod. Phys. {\bf 81}, 299 (2009)}}.
	
	\bibitem{Manko1999}
	V.I. Man’ko and R.V. Mendes, 
	Non-commutative time-frequency tomography of analytic signals, 
	{\href{http://dx.doi.org/10.1016/S0375-9601(99)00688-X}{Phys. Lett. A {\bf 263}, 53 (1999)}}.
	
	\bibitem{MankoMA2006}
	M.A. Man'ko,
	Entropy of an analytic signal,
	{\href{http://dx.doi.org/10.1007/s10946-006-0023-y}{J. Russ. Laser Res. {\bf 21}, 411 (2000)}}.
	
	\bibitem{Mendes2001}
	M.A. Man'ko, V.I. Man’ko, and R.V. Mendes,
	Tomograms and other transforms: a unified view,
	{\href{http://dx.doi.org/10.1088/0305-4470/34/40/309}{J. Phys. A: Math. Gen. {\bf 34}, 8321 (2001)}}.
	
	\bibitem{Manko2005}
	S. De Nicola, R. Fedele, M.A. Man'ko and V.I. Man'ko, 
	Fresnel tomography: A novel approach to wave-function reconstruction based on the Fresnel representation of tomograms, 
	{\href{http://dx.doi.org/10.1007/s11232-005-0151-5}{Theor. Math. Phys. {\bf 144}, 1206 (2005)}}.
	
	\bibitem{MankoMA2000}
	M.A. Man'ko,
	Quasidistributions, tomography, and fractional Fourier transform in signal analysis,
	{\href{http://dx.doi.org/10.1007/BF02508735}{J. Russ. Laser Res. {\bf 21}, 411 (2000)}}.
	
	\bibitem{Manko2012}
	F. Briolle, V. I. Man'ko, B. Ricaud, R.V. Mendes, 
	Non-commutative tomography: A tool for data analysis and signal processing, 
	{\href{http://dx.doi.org/10.1007/s10946-012-9265-z}{J. Russ. Laser Res. {\bf 33}, 103 (2012)}}.
	
	\bibitem{Mendes2009}
	F. Briolle, R. Lima, V.I. Man'ko, and R.V. Mendes,
	A tomographic analysis of reflectometry data: I. Component factorization,
	{\href{http://dx.doi.org/10.1088/0957-0233/20/10/105501}{Meas. Sci. Technol. {\bf 20}, 105501 (2009)}}.	
	
	\bibitem{Mendes2015}
	R.V. Mendes,
	Non-commutative tomography and signal processing,
	{\href{http://dx.doi.org/10.1007/s11232-005-0151-5}{Phys. Scr. {\bf 90}, 074022 (2015)}}.
	
	\bibitem{Trajin2015}
	B. Trajin, M. Chabert, J. Regnier, and J. Faucher,
	Wigner distribution for the diagnosis of high frequency amplitude and phase modulations on stator currents of induction machine, in
	{\href{https://doi.org/10.1109/DEMPED.2009.5292804}{{\it Proceedings of IEEE International Symposium on Diagnostics for Electric Machines, Power Electronics and Drives}, Cargese, France (IEEE, New York, 2009)}}.
	
	\bibitem{Shin1993}
	Y. S. Shin and J.-J. Jeon,
	Pseudo Wigner-Ville time-frequency distribution and its application to machinery condition monitoring,
	{\href{https://doi.org/10.3233/SAV-1993-1109}{Shock Vib. {\bf 1}, 65 (1993)}}.

	\bibitem{Katkovnik1998}	
	V. Katkovnik and L. Stankovi\'{c},
	Instantaneous frequency estimation using the Wigner distribution with varying and data-driven window length,
	{\href {https://doi.org/10.1109/78.709514}{IEEE Trans. Signal Process. {\bf 46}, 2315 (1998)}}.

	\bibitem{Katkovnik19982}
	L. Stankovi\'{c} and V. Katkovnik,
	Algorithm for the instantaneous frequency estimation using time-frequency distributions with adaptive window width,
	{\href{https://doi.org/10.1109/LSP.2009.2027651}{IEEE Signal Process. Lett. {\bf 5}, 224 (1998)}}.
	
	\bibitem{Sucic2009}
	J. Lerga and V. Sucic,
	Nonlinear IF estimation based on the pseudo WVD adapted using the improved sliding pairwise ICI rule,
	{\href {https://doi.org/10.1109/LSP.2009.2027651}{IEEE Signal Process. Lett. {\bf 16}, 953 (2009)}}.
	
	\bibitem{Jukam2016}
	T. Fobbe, S. Markmann, F. Fobbe, N. Hekmat, H. Nong, S. Pal, P. Balzerwoski, J. Savolainen, M. Havenith, A. D. Wieck, and N. Jukam,
	Broadband terahertz dispersion control in hybrid waveguides,
	{\href{https://doi.org/10.1364/OE.24.022319}{Opt. Express {\bf 24}, 22319 (2016)}}.
	
	\bibitem{Jin2016}
	Z.-K. Gao, S.-S. Zhang, Q. Cai, Y.-X. Yang and N.-D. Jin
	Complex network analysis of phase dynamics underlying oil-water two-phase flows,
	{\href{https://doi.org/10.1038/srep28151}{Sci. Rep. {\bf 6}, 28151 (2016)}}.

	\bibitem{Nishad2016}	
	R.B. Pachori and A. Nishad,
	Cross-terms reduction in the Wigner-Ville distribution using tunable-Q wavelet transform,
	{\href{https://doi.org/10.1016/j.sigpro.2015.07.07.026}{Signal Process. {\bf 120}, 288 (2016)}}.
	
	\bibitem{Mustard2006}
	D. Mustard,
	The fractional Fourier transform and the Wigner distribution,
	{\href{https://doi.org/10.1017/S0334270000000606} {J. Austral. Math. Soc. Ser. B {\bf 38}, 209 (1996)}}.
	
	\bibitem{Lima2017}
	J.R. de Oliveira Neto and J.B. Lima,
	Discrete fractional Fourier transforms based on closed-form Hermite-Gaussian-like DFT eigenvectors,
	{\href{https://doi.org/10.1109/TSP.2017.2750105}{IEEE Trans. Signal Process. {\bf 65}, 6171 (2017)}}.
	
	\bibitem{Mycielski1975}
	I. Biatynicki-Birula and J. Mycielski,
	Uncertainty relations for information entropy in wave mechanics,
	{\href{https://doi.org/10.1007/BF01608825}{Commun. Math. Phys. {\bf 44}, 129 (1975)}}.

\end{thebibliography}
\end{document}